\documentclass[11pt]{article}
\usepackage{authblk}
\usepackage[bookmarksnumbered, colorlinks, plainpages]{hyperref}
\usepackage{amsmath, amsthm, amscd, amsfonts, amssymb, graphicx, color, booktabs}
\numberwithin{equation}{section}
\definecolor{email}{rgb}{0.00,0.00,0.84}

\def\sigmat{\sigma^{}_t}
\def\simge{\mathrel{%
   \rlap{\raise 0.511ex \hbox{$>$}}{\lower 0.511ex \hbox{$\sim$}}}}
\def\simle{\mathrel{
   \rlap{\raise 0.511ex \hbox{$<$}}{\lower 0.511ex \hbox{$\sim$}}}}

\begin{document}
\setcounter{page}{1}

\title{\large \bf 12th Workshop on the CKM Unitarity Triangle \\ 
Santiago de Compostela, 18-22 September 2023 \\ 
\vspace{0.3cm}
\LARGE Charm lifetime measurements at Belle~II}

\author{Alan Schwartz\textsuperscript{1} on behalf of the Belle~II Collaboration \\
        \textsuperscript{1}Physics Department, University of Cincinnati, Cincinnati, Ohio 45221\\ }
\maketitle

\begin{abstract}
We report precision measurements of lifetimes of charmed mesons and baryons performed by the
Belle~II experiment. Specifically, we measure $D_s^+$, $D^+$, $D^0$, $\Lambda_c^+$, and $\Omega_c^0$
lifetimes. Our results for $\tau(D_s^+)$, $\tau(D^+)$, $\tau(D^0)$, and $\tau(\Lambda_c^+)$ 
are the world's most precise; our result for $\tau(\Omega_c^0)$ confirms that the $\Omega_c^0$
lifetime is longer than that of the $\Lambda_c^+$ and the $\Xi_c^0$.
\end{abstract} 

\maketitle

\section{Introduction}

The lifetime of a particle, like its mass, is one of its fundamental properties. 
The lifetime is the reciprocal of the sum of all partial widths, and thus all 
final states contribute to it. In this manner, the lifetime can provide 
information about final states that are difficult to measure or calculate.

Charm lifetimes are calculated within the framework of the Heavy Quark Expansion (HQE)~\cite{Lenz}. 
Using the optical theorem, the expansion takes the form (for a $D$ meson)~\cite{Kingetal}:
\begin{eqnarray}
\Gamma ({D})  & = & 
\frac{1}{2 m_{{D}}} \sum_{X}  \int \limits_{\rm PS} (2 \pi)^4  \delta^{(4)}(p^{}_{D}- p^{}_X) \, 
|\langle X(p^{}_X)| {\cal H}_{\rm eff} | {D}(p^{}_{D}) \rangle |^2 \label{eq:Gamma-D} 
\label{eqn:width1} \\
 & \rightarrow & 
\Gamma_3  +
\Gamma_5 \frac{\langle {\cal O}_5 \rangle}{m_c^2} + 
\Gamma_6 \frac{\langle {\cal O}_6 \rangle}{m_c^3} + ...  
 + 16 \pi^2 
\left( 
  \tilde{\Gamma}_6 \frac{\langle \tilde{\mathcal{O}}_6 \rangle}{m_c^3} 
+ \tilde{\Gamma}_7 \frac{\langle \tilde{\mathcal{O}}_7 \rangle}{m_c^4} + \ldots \right) ,
\label{eqn:width2} 
\end{eqnarray}
where the summation in Eq.~(\ref{eqn:width1}) is over all final states, and 
the $\Gamma_n$ terms in Eq.~(\ref{eqn:width2}) are Wilson coefficients that 
are expanded in powers of $\alpha^{}_s$ and calculated perturbatively. 
The $\langle \mathcal{O}_n \rangle$ and $\langle \tilde{\mathcal{O}}_n \rangle$ terms
are matrix elements of dimension-$n$ local operators and must be calculated using 
non-perturbative methods.\footnote{No tilde denotes a two-quark operator; 
a tilde denotes a four-quark operator~\cite{Kingetal}.}
Comparing HQE calculations with experimental measurements tests our 
understanding of QCD. Here we present several lifetime measurements
of charmed hadrons performed by the Belle~II experiment~\cite{BelleII_web}.

Belle~II runs at the SuperKEKB $e^+e^-$ collider~\cite{SuperKEKB} based at the KEK laboratory 
in Tsukuba, Japan. The experiment studies weak decays of $B$ and $D$ mesons, and $\tau$ 
leptons, with the goal of uncovering new physics. The Belle~II detector is described  
in Ref.~\cite{BelleII_detector}. The measurements reported here were performed with early 
data sets, before the second layer of the silicon pixel detector was fully installed. 
The lifetime is determined from a fit to the decay time distribution. For a particle 
of mass $m$ and momentum $\vec{p}$, the decay time is calculated as 
$t = m\,(\vec{d}\cdot\hat{p})/|\vec{p}\,|$,
%\begin{eqnarray}
%t & = & \left(\frac{\vec{d}\cdot\hat{p}}{p}\right) m^{}_{D} \,,
%\end{eqnarray}
where $\vec{d}$ is the displacement vector from the $e^+e^-$ interaction point (IP), 
where the particle is produced, to its decay vertex position. 
%$\vec{p}$ is its measured momentum; and $m^{}_{D}$ is the $D$ mass. 
The IP in Belle~II is measured
every 30 minutes using $e^+e^-\to\mu^+\mu^-$ events. All charm lifetime analyses 
impose a momentum requirement on the decaying meson or baryon to eliminate those 
originating from $B$ decays, which are displaced from the IP and would corrupt 
the lifetime measurement.

The uncertainty on $t$, denoted $\sigma^{}_t$, is calculated event-by-event by 
propagating the uncertainties on $\vec{d}$ and $\vec{p}$, taking into account their 
correlations. The decay time resolution ($\langle \sigma^{}_t\rangle$) is approximately 
twice as precise as that achieved at Belle and Babar: $80\!-\!90$~fs versus $\sim\!200$~fs. 
For the analyses presented here, the uncertainty $\sigma^{}_t$ multiplied by a scaling
factor is taken as the width of a Gaussian resolution function used to fit the decay 
time distributions.

\section{\boldmath $D_s^+$ lifetime}

The most recent measurement is that of the $D_s^+$ lifetime, which used 207~fb$^{-1}$ of 
data~\cite{belleII_Ds}. The Cabibbo-favored (CF) decay $D_s^+\to\phi (\to K^+K^-) \pi^+$ 
is reconstructed, and candidates with an invariant mass satisfying
$M(\phi\pi^+)\in [1.960,1.976]$~GeV/$c^2$ are used to measure the lifetime. Fitting 
the $M(\phi\pi^+)$ distribution, we obtain a signal yield in this range of 
$116\times 10^3$ events with a purity of~92\%. 

The lifetime is determined from an unbinned maximum likelihood fit to the decay time
($t$) distribution. As the uncertainty $\sigma^{}_t$ is used in the resolution function, 
and its distribution differs for signal and background events, we include probability 
density functions (PDFs) for $\sigma^{}_t$ in the likelihood function to avoid
biasing the fit results~\cite{Punzi}. The likelihood function for event $i$ 
is given by
\begin{eqnarray}
{\cal L}(\tau | t^i,\sigmat{}\!\!^i) & = &   
f_{\rm sig}\,P_{\rm{sig}}(t^i | \tau, \sigmat{}\!\!^i)\,P_{\rm{sig}}(\sigmat{}\!\!^i)\ +\   
(1-f_{\rm{sig}})\,P_{\rm{bkg}}(t^i|\sigmat{}\!\!^i)\,P_{\rm{bkg}}(\sigmat{}\!\!^i) \,,
\end{eqnarray}
where $f^{}_{\rm sig}$ is the fraction of events that are signal decays, and 
$P_{\rm{sig}}(t^i|\tau,\sigma_t^i)$ is the signal PDF for measuring $t^{}_i$ 
given a lifetime $\tau$ and an uncertainty $\sigma_t^i$. 
The background PDF $P^{}_{\rm bkg}(t^i|\sigmat{}\!\!^i)$ is an exponental 
function determined from events in the sideband $M(\phi\pi^+)\in [1.990,2.020]$~GeV/$c^2$.
The PDF $P_{\rm{bkg}}(\sigmat{}\!\!^i)$ is a histogram also determined from the 
$M(\phi\pi^+)$ sideband, and $P_{\rm{sig}}(\sigmat{}\!\!^i)$ is a histogram determined 
from events in the $M(\phi\pi^+)$ signal region after subtracting off the 
$P_{\rm{bkg}}(\sigmat{}\!\!^i)$ contribution. The fraction $f^{}_{\rm sig}$ 
is determined from the fit to the $M(\phi\pi^+)$ distribution.

\begin{figure} [h!]
\hspace*{-0.20in}
\includegraphics[width=230pt]{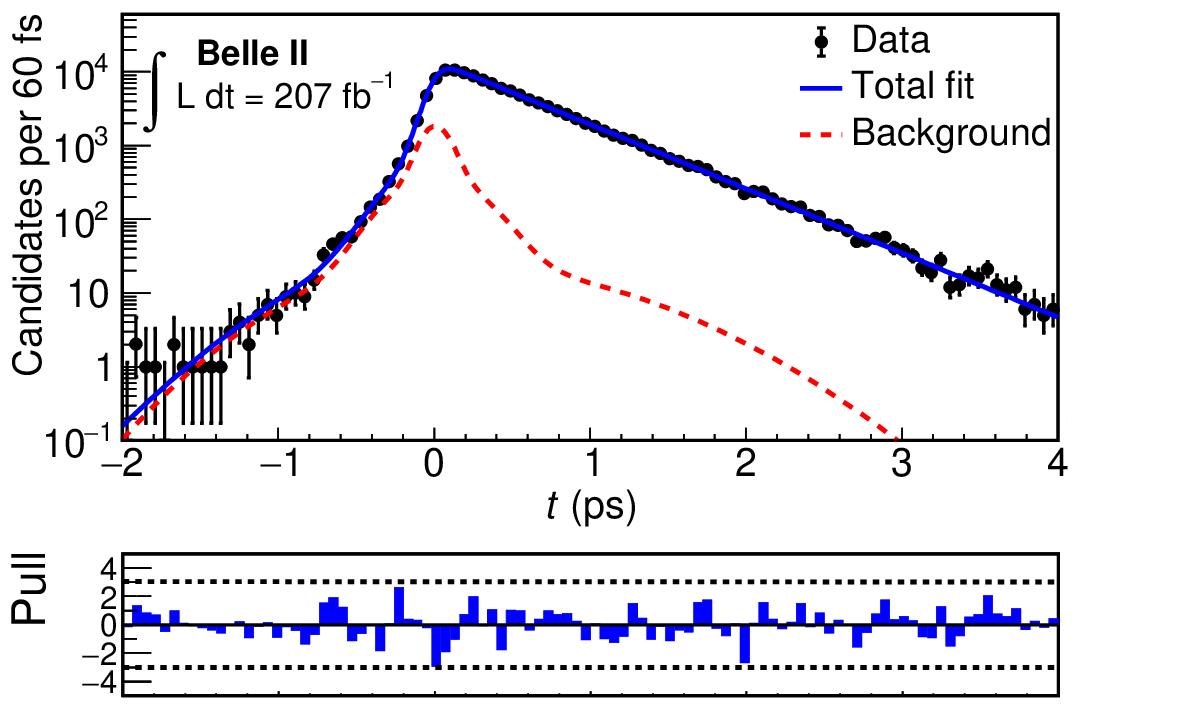}
\hspace*{-0.15in}
\includegraphics[width=230pt]{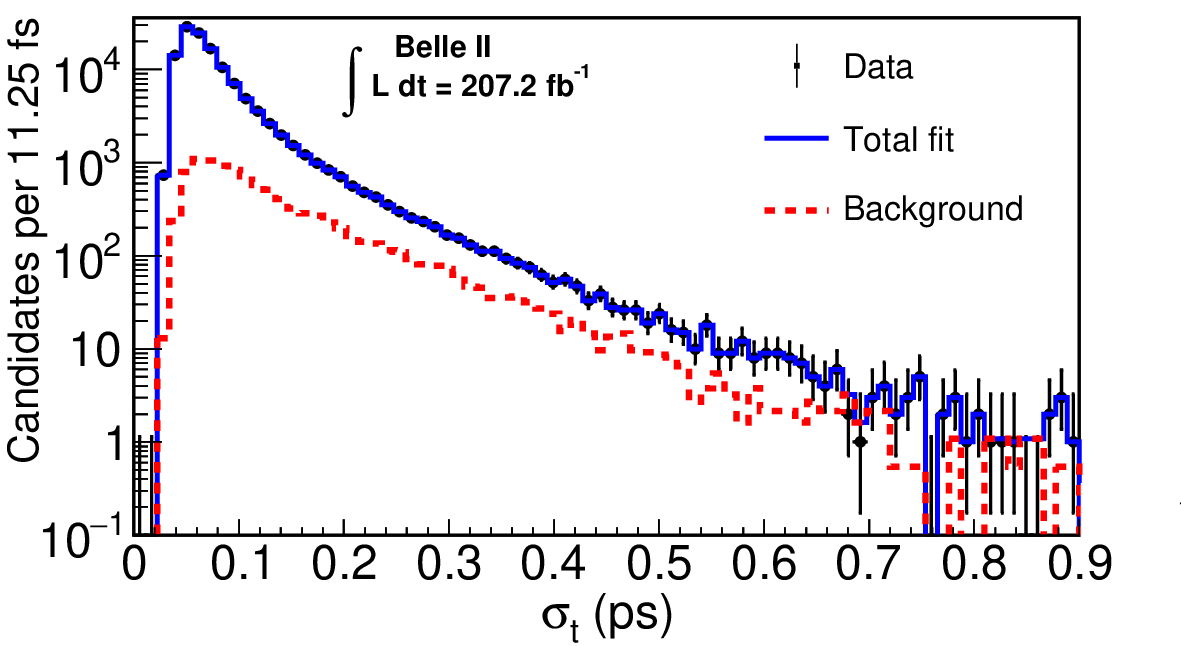}
\caption{$D_s^+\to\phi\pi^+$ decay time distribution with the fit result superimposed (left),
and $\sigma_t$ distributions for background and signal\,+\,background (right)~\cite{belleII_Ds}.}
\label{fig:fit_result_Ds}.
\end{figure}

The PDF for signal decays is
\begin{eqnarray}
P^{}_{\rm sig}(t^i|\tau, \sigmat{}\!\!^i)  
\ =\ \frac{1}{\tau} \int e^{-t'/\tau}\,R(t^i-t' ; \mu, s, \sigmat{}\!\!^i)\,dt' \,,
\end{eqnarray}
where $R$ is a Gaussian function with mean $\mu$ and a per-candidate standard 
deviation $s\cdot\sigma_t^i$, with $s$ being a scaling factor. The lifetime $\tau$ 
is determined by maximizing the total likelihood $\Sigma_i {\cal L}_i$, where 
the summation runs over all signal candidates.
%fitted $M(\phi\pi^+)$ region. 
The floated parameters are $\tau$, $\mu$, and~$s$. The result of the fit is 
$\tau(D^+_s) \ =\ (499.5 \pm 1.7 \pm 0.9)$~fs, where the first uncertainty is
statistical and the second is systematic. 
The projection of the fit result is shown in Fig.~\ref{fig:fit_result_Ds}.
The sources of systematic uncertainty are listed in Table~\ref{tab:syst_Ds}. 

\begin{table}[h]
\centering
\caption{Systematic uncertainties for the $D_s^+$ lifetime measurement~\cite{belleII_Ds}.}
\vskip0.10in
\begin{tabular}{lc} 
\hline 
Source                                   & Uncertainty (fs)   \\ 
\hline    
Resolution function                      &   $+ 0.43$       \\
Background $(t,\sigma^{}_t)$ distribution  &  $\pm 0.40$       \\
Binning of $\sigmat$ histogram PDF      &   $\pm 0.10$       \\
Imperfect detector alignment             &  $\pm 0.56$       \\
Sample purity                            &   $\pm 0.09$       \\
Momentum scale factor                    &   $\pm 0.28$       \\
$D^+_s$ mass                             &   $\pm 0.02$       \\ \hline
Total                                    &   $\pm 0.87$ \\ 
\hline 
\end{tabular}
\label{tab:syst_Ds}
\end{table}

\section{\boldmath $D^+$ and $D^0$ lifetimes}

The measurements of the $D^0$ and $D^+$ lifetimes~\cite{belleII_D0Dplus} are similar 
to that for the $D_s^+$ but use a smaller data set: 78~fb$^{-1}$. To reduce backgrounds, 
$D^0$ and $D^+$ candidates are required to originate from either $D^{*+}\to D^0\pi^+_s$ 
or $D^{*+}\to D^+\pi^0_s$ decays. The pion daughter in these decays has very low 
momentum in the laboratory frame and thus is labeled the ``slow'' pion ($\pi^{}_s$).
The $D^0$ and $D^+$ are reconstructed in the CF final states $K^-\pi^+$ and 
$K^-\pi^+\pi^+$, respectively. To eliminate signal candidates originating 
from $B\to DX$, we require that the $D^*$ momentum in the $e^+e^-$ 
center-of-mass (CM) frame be greater than 2.5~GeV/$c$ for $D^0$ candidates 
and $>\!2.6$~GeV/$c$ for $D^+$ candidates. 
For the lifetime fits, we select $D^0$ candidates satisfying
$M(K^-\pi^+)\in[1.851,1.878]$~GeV/$c^2$ and 
$M(K^-\pi^+\pi^+_s)\!-\!M(K^-\pi^+) \in [144.94,145.90]$~MeV/$c^2$,
and $D^+$ candidates satisfying 
$M(K^-\pi^+\pi^+)\in[1.855,1.883]$~GeV/$c^2$ and 
$M(K^-\pi^+\pi^+\pi^0_s)\!-\!M(K^-\pi^+\pi^+) \in [138.0,143.0]$~MeV/$c^2$.
These samples contain 
$171\times 10^3$ $D^0\to K^-\pi^+$ candidates (99.8\% purity) and
$59\times 10^3$ $D^0\to K^-\pi^+\pi^+$ candidates (91\% purity). 

\begin{figure} [h!]
\begin{center}
\includegraphics[width=250pt]{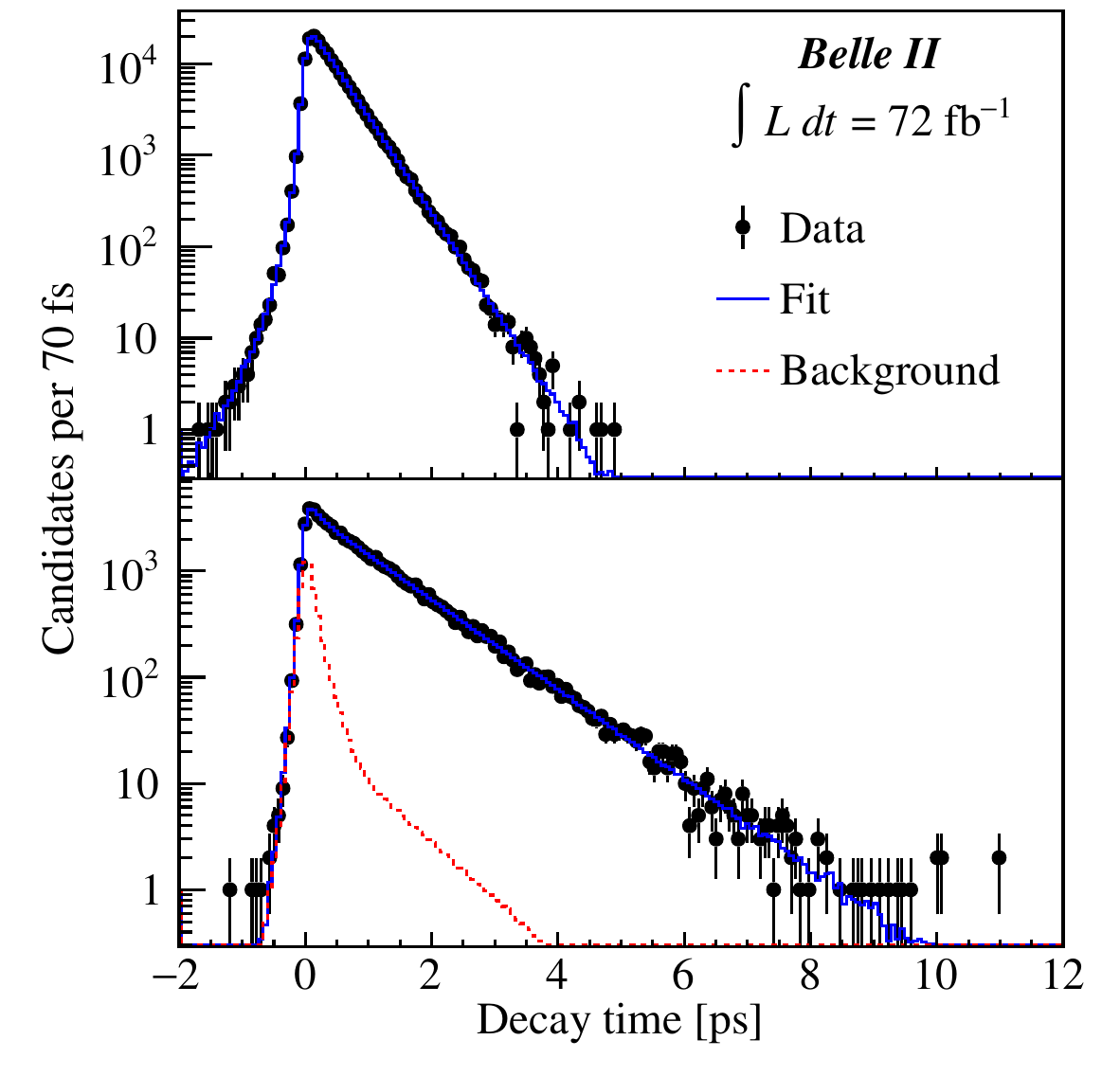}
\end{center}
\vspace*{-0.30in}
\caption{Decay time distribution with the fit result superimposed
for $D^0\to K^-\pi^+$ (top) and $D^+\to K^-\pi^+\pi^+$ (bottom)~\cite{belleII_D0Dplus}.}
\label{fig:fit_results_D}.
\end{figure}

As in the $D_s^+$ analysis, the lifetimes are determined from unbinned 
maximum likelihood fits to the decay time, and the per-candidate
uncertainty $\sigma_t$ multiplied by a scaling factor is used as the width(s)
of a double Gaussian ($D^0$) or single Gaussian ($D^+$) resolution function.
The decay time PDF for $D^+$ background is obtained from the sideband 
$M(K^-\pi^+\pi^+)\in[1.758,1.814] \cup [1.936,1.992]$~GeV/$c^2$,
whereas $D^0$ background is negligible.
The results of the fits are
$\tau(D^0) = (410.5 \pm 1.1\,\pm 0.8)$~fs and
$\tau(D^+) = (1030.4 \pm 4.7\,\pm 3.1)$~fs,  
where the first uncertainty is statistical and the second is systematic.
The decay time distributions are shown in Fig.~\ref{fig:fit_results_D} 
along with projections of the fit result. The sources of systematic 
uncertainty are listed in Table~\ref{tab:systs_D}.

\begin{table}[h]
\centering
\caption{Systematic uncertainties for the $D^0$ and $D^+$ lifetime measurements~\cite{belleII_D0Dplus}.}
\vskip0.10in
\begin{tabular}{lcc}
\hline
Source & $\tau(D^0)$ (fs) & $\tau(D^+)$ (fs) \\
\hline
Resolution model   & 0.16 & 0.39 \\
Backgrounds        & 0.24 & 2.52 \\
Detector alignment & 0.72 & 1.70 \\
Momentum scale     & 0.19 & 0.48 \\ \hline
Total              & 0.80 & 3.10 \\
\hline
\end{tabular}
\label{tab:systs_D}
\end{table}

\section{\boldmath $\Lambda_c^+$ lifetime}

Measurement of the $\Lambda_c^+$ lifetime~\cite{belleII_Lambdac} is similar to 
that for the $D_{(s)}^+$ and $D^0$ lifetimes. The reconstructed decay is 
$\Lambda_c^+\to p K^-\pi^+$. However, some $\Lambda_c^+$ candidates 
originate from $\Xi_c^0\to \Lambda_c^+\pi^-$ and 
$\Xi_c^+\to \Lambda_c^+\pi^0$ decays; since $\tau(\Xi_c^+)=152$~fs and 
$\tau(\Xi_c^0)=453$~fs~\cite{PDG24}, such $\Lambda_c^+$ candidates 
corrupt the $\tau(\Lambda_c^+)$ measurement. Fortunately, they can be 
identified from the displacement of the $\Lambda_c^+$ decay vertex
from the IP in the plane transverse to the beam line: $\Lambda_c^+$ 
candidates from $\Xi_c$ decays typically have larger values. Fitting 
this distribution yields $374\pm 88$ $\Xi_c\to\Lambda_c^+ X$ decays in the 
entire $\Lambda_c^+$ sample. These events are reduced by 40\% by 
combining the $\Lambda_c^+$ candidate with a $\pi^-$ or $\pi^0$ 
candidate and vetoing those that satisfy
$M(pK^-\pi^+\pi^-) - M(p K^-\pi^+)\in [183.4,186.4]$~MeV/$c^2$ or 
$M(pK^-\pi^+\pi^0) - M(p K^-\pi^+)\in [175.3,187.3]$~MeV/$c^2$.
The effect of the remaining $\Lambda_c^+$ candidates is 
evaluated using MC simulation; the resulting bias of $+0.34$~fs is 
subtracted from the fitted $\Lambda_c^+$ lifetime and included 
as a systematic uncertainty (see Table~\ref{tab:syst_Lambdac}). The 
final fitted sample consists of $116\times 10^3$ events with 
92.5\% purity. The fit result after the bias correction is
$\tau(\Lambda_c^+) = (203.20\pm 0.89\pm 0.77)$~fs.

\begin{table}[h]
\centering
\caption{Systematic uncertainties for the $\Lambda_c^+$ lifetime measurement~\cite{belleII_Lambdac}.}
\vskip0.10in
\begin{tabular}{lc}
\hline
Source & Uncertainty [fs] \\
\hline
Resolution model          & 0.46 \\
$\Xi_{c}$ contamination   & 0.34 \\
Non-$\Xi_{c}$ backgrounds & 0.20 \\
Detector alignment        & 0.46 \\
Momentum scale            & 0.09 \\
\hline
Total                     & 0.77 \\
\hline
\end{tabular}
\label{tab:syst_Lambdac}
\end{table}

\section{\boldmath $\Omega_c^0$ lifetime} 

\begin{figure} [h!]
\hbox{
\vbox{
\hspace*{-0.40in}
\vspace*{0.40in}
\includegraphics[width=230pt]{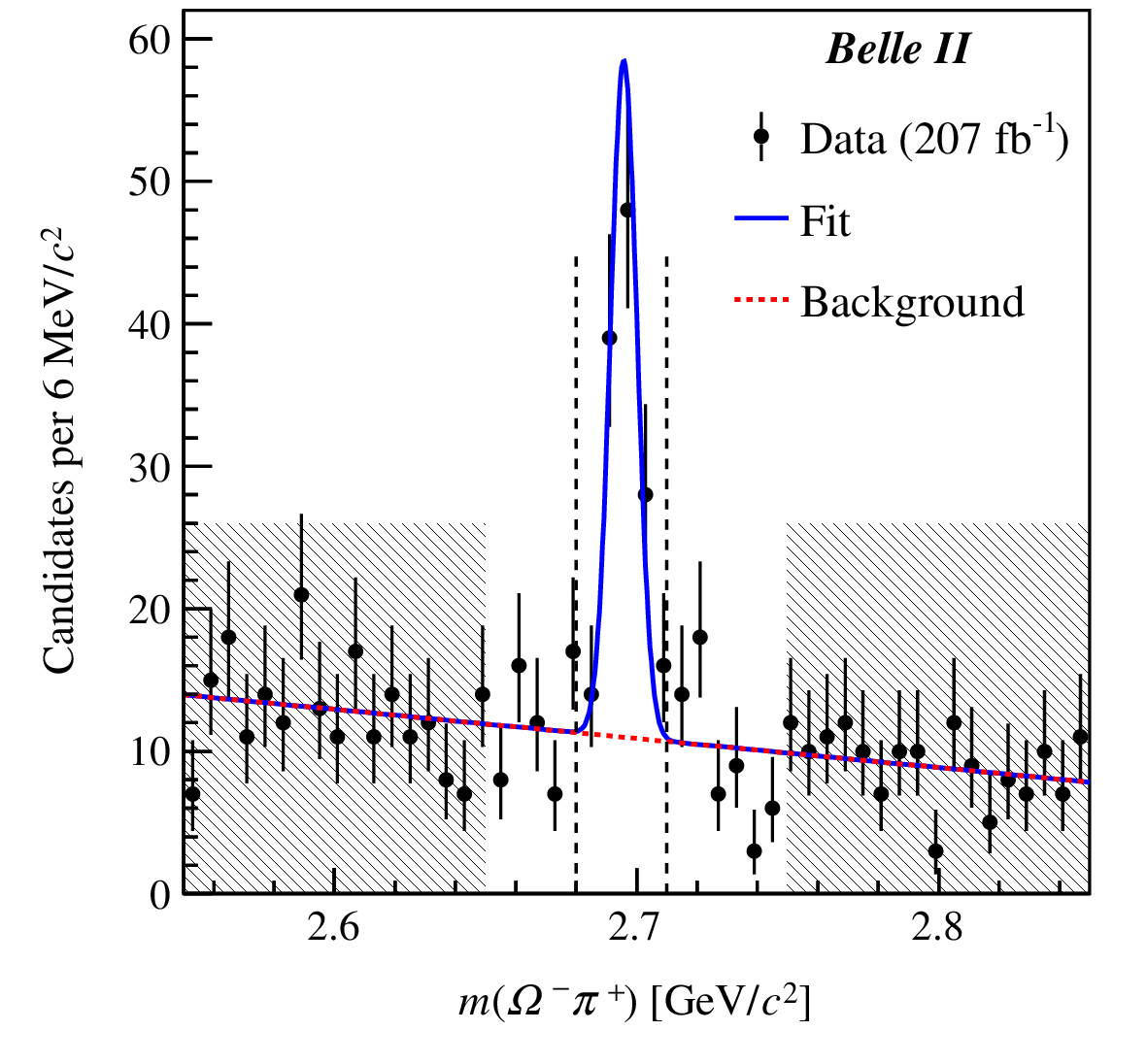}}
%\hspace*{-0.15in}
\hspace*{-3.1in}
\includegraphics[width=230pt]{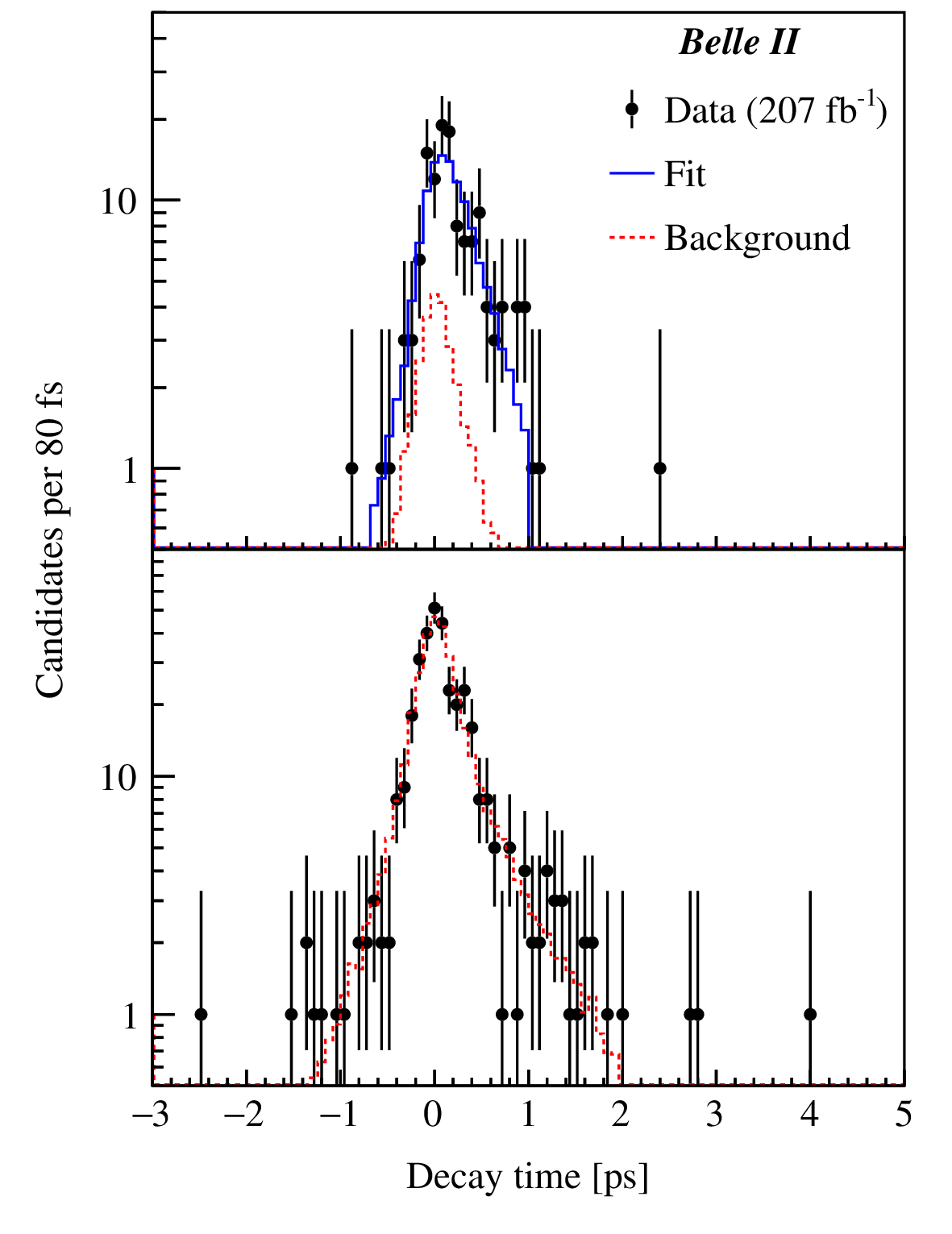}}
\vspace*{-0.15in}
\caption{Invariant mass distribution for $\Omega_c^0\to\Omega^- (\to \Lambda K^-)\pi^+$
candidates (left);
decay time distribution for $\Omega_c^0$ candidates in the signal region
$M(\Omega^-\pi^+)\in[2.68,2.71]$~GeV/$c^2$ (right top); 
decay time distribution for $\Omega_c^0$ candidates in the sideband
$M(\Omega^-\pi^+)\in[2.55,2.65] \cup [2.75,2.85]$~GeV/$c^2$ (right bottom).
The latter events are fitted to obtain the decay time PDF used for 
background in the signal region~\cite{belleII_Omegac}.}
\label{fig:fit_Omegac}.
\end{figure}

Measurement of the $\Omega_c^0$ lifetime~\cite{belleII_Omegac} follows the same 
procedure as described above for the $\Lambda_c^+, D^+_{(s)}$, and $D^0$ lifetimes.
The reconstructed decay is 
$\Omega_c^0\to \Omega^- \pi^+$ followed by $\Omega^-\to\Lambda (\to p\pi^-) K^-$. 
To eliminate $\Omega_c^0$ candidates originating from $B\to\Omega_c^0 X$,
we require $p^{}_{\Omega_c^0}/p^{}_{\rm max} > 0.60$, where
$p^{}_{\rm max}=\sqrt{(E^{}_{\rm beam})^2 - m(\Omega^-\pi^+)^2}$, and
the momentum $p^{}_{\Omega^0_c}$ and beam energy $E^{}_{\rm beam}$ are
evaluated in the $e^+e^-$ CM frame. The final candidate sample
is much smaller than those of the other lifetime measurements: 
132 signal candidates with 67\% purity -- see Fig.~\ref{fig:fit_Omegac}~(left).
The final result is $\tau(\Omega_c^0) = (243\pm 48\pm 11)$~fs, where the
first uncertainty is statistical and the second is systematic. The
systematic uncertainty is dominated by the resolution function ($\pm 6.2$~fs)
and background modeling ($\pm 8.3$)~fs. The fitted decay time distribution is 
shown in Fig.~\ref{fig:fit_Omegac}~(right) along with projections of the fit result.
The measured lifetime is significantly longer than the lifetime measured by
Fermilab E687~\cite{E687_Omegac}, CERN WA89~\cite{WA89_Omegac}, and FOCUS~\cite{FOCUS_Omegac}, 
but it agrees with more recent measurements by LHCb~\cite{LHCb_Omegac}. This lifetime 
is unexpectedly longer than that of lighter charmed baryons, i.e., the hierarchy of 
charm baryon lifetimes becomes 
$\tau(\Xi_c^+) > \tau(\Omega_c^0) > \tau(\Lambda_c^+) > \tau(\Xi_c^0)$.

\section{Summary}

Except for the $\Omega_c^0$ charm baryon, the lifetime measurements presented here are 
the world's most precise. For the $\Omega_c^0$, the Belle~II measurement confirms the 
unexpectedly long lifetime measured by LHCb~\cite{LHCb_Omegac}. 
All measurements are compared to theory predictions~\cite{Kingetal,Gratrexetal} 
in Table~\ref{tab:theory_exp_comp}.
Currently, the theory uncertainties are large and preclude a precision test of QCD.
While all measurements agree with theory predictions within~$2\sigma$, 
the measured values of $\tau(D^0)$, $\tau(D^+)$, and $\tau(\Lambda_c^+)$ 
differ from the predictions by more than~$1\sigma$. Such differences can 
help discriminate among different calculational approaches and ultimately 
improve our understanding of QCD.

\begin{table}[h]
\centering
\caption{Comparison between measurements and HQE theory predictions for charm lifetimes.
The limits listed correspond to 90\% C.L. The HQE calculations for $\tau(D_s^+)$ do
not include the partial width for $D_s^+\to\tau^+\nu$, as $m^{}_\tau > m^{}_c$.}
\renewcommand{\arraystretch}{1.25}
\vskip0.10in
\begin{tabular}{l|ccc}
\hline
        & Belle~II & King et al.~\cite{Kingetal} & Gratrex et al.~\cite{Gratrexetal} \\
        &          &                             &  (MSR scheme)                     \\
\hline
$\tau(D^0)$ [fs]   & $410.5\pm 1.1\pm 0.8$~\cite{belleII_D0Dplus}  
                              & $629\,^{+296}_{-167}$ & $595\,^{+344}_{-166}$  \\
$\tau(D^+)$ [fs]   & $1030.4\pm 4.7\pm 3.1$~\cite{belleII_D0Dplus}   & $>1193$  & $>1690$   \\
$\tau(D^+_s)$ [fs] & $499.5\pm 1.7\pm 0.9$~\cite{belleII_Ds}   & $637\,^{+381}_{-190}$ & $599\,^{+459}_{-180}$  \\
$\tau(D^+)/\tau(D^0))$  & $2.510\pm 0.016$   & $2.80\,^{+0.86}_{-0.90}$   & $2.89\,^{+0.78}_{-0.85}$   \\
$\tau(D^+_s)/\tau(D^0)$ & $1.217\pm 0.006$   & $1.01\pm 0.15$   & $1.00\,^{+0.23}_{-0.21}$   \\
$\tau(\Lambda_c^+)$ [fs] & $203.20\pm 0.89\pm 0.77$~\cite{belleII_Lambdac}    &    &  $312\,^{+128}_{-96}$  \\
$\tau(\Omega_c^0)$  [fs] & $243\pm 48\pm 11$~\cite{belleII_Omegac}   &    &  $237\,^{+111}_{-75}$  \\
$\tau(\Omega_c^0)/\tau(\Lambda_c^+)$ & $1.20\pm 0.24$    &    &  $0.83\,^{+0.30}_{-0.18}$  \\
\hline
\end{tabular}
\label{tab:theory_exp_comp}
\end{table}

\bibliographystyle{amsplain}

\end{document}